\documentclass[%
 reprint,
superscriptaddress,
 amsmath,amssymb,
 aps,
prb,
]{revtex4-2}

\usepackage{graphicx}
\usepackage{dcolumn}
\usepackage{bm}

\begin{document}


\title{Side-gate modulation of supercurrent in InSb nanoflag-based Jo\-seph\-son junctions}

\author{Bianca Turini}
\affiliation{NEST, Istituto Nanoscienze-CNR and Scuola Normale Superiore, Piazza San Silvestro 12, 56127 Pisa, Italy}

\author{Sedighe Salimian}
\affiliation{NEST, Istituto Nanoscienze-CNR and Scuola Normale Superiore, Piazza San Silvestro 12, 56127 Pisa, Italy}

\author{Matteo Carrega}
\email{matteo.carrega@spin.cnr.it}
\affiliation{CNR-SPIN, Via Dodecaneso 33, 16146 Genova, Italy}

\author{Federico Paolucci}
\altaffiliation[Present address: ]{Dipartimento di Fisica "E. Fermi", Università di Pisa, Largo Pontecorvo 3, I-56127 Pisa, Italy}
\affiliation{NEST, Istituto Nanoscienze-CNR and Scuola Normale Superiore, Piazza San Silvestro 12, 56127 Pisa, Italy}

\author{Valentina Zannier}
\affiliation{NEST, Istituto Nanoscienze-CNR and Scuola Normale Superiore, Piazza San Silvestro 12, 56127 Pisa, Italy}

\author{Lucia Sorba}
\affiliation{NEST, Istituto Nanoscienze-CNR and Scuola Normale Superiore, Piazza San Silvestro 12, 56127 Pisa, Italy}

\author{Stefan Heun}
\email{stefan.heun@nano.cnr.it}
\affiliation{NEST, Istituto Nanoscienze-CNR and Scuola Normale Superiore, Piazza San Silvestro 12, 56127 Pisa, Italy}

\date{\today}

\begin{abstract}
InSb nanoflags, due to their intrinsic spin-orbit interactions, are an interesting platform in the study of planar Josephson junctions. Ballistic transport, combined with high transparency of the superconductor/semiconductor interfaces, was reported to lead to interesting phenomena such as the Josephson diode effect. The versatility offered by the planar geometry can be exploited to manipulate both carrier concentration and spin-orbit  strength by electrical means. Here we present experimental results on InSb nanoflag-based Josephson junctions fabricated with side-gates placed in close proximity to the junction. We show that side-gates can efficiently modulate the current through the junction, both in the dissipative and in the dissipation-less regimes, similarly to what obtained with a conventional back-gate. Furthermore, the side-gates can be used to influence the Fraunhofer interference pattern induced by the presence of an external out-of-plane magnetic field.
\end{abstract}

\maketitle


\section{Introduction}

Hybrid superconductor/semiconductor devices have attracted a lot of interest in the last decades due to their potential applications in a wide range of fields, such as radiation sensing, quantum computing, particle detection, and telecommunications~\cite{Moehle2021, Bercioux2015, Manchon2015, Dayton2018}. Among all, the pursuit for new topological states of matter~\cite{Sato2017,Lutchyn2018,Banerjee2023,Tanaka2024,Schiela2024,Dartiailh2021} triggered intense research on strong spin-orbit semiconductors interfaced with superconductors. Progress in fabrication techniques has been achieved recently~\cite{Moehle2021, Vaitiekenas2018, Matsuo2023, Ciaccia2023, Verma2021, Ke2019, Vries2019, Salimian2021, Erlandsson2023, Banerjee2023a, Danilenko2023}, as well as more efficient models to describe proximity effects~\cite{Kulik1970, Pannetier2000, Brammertz2001, Bespalov2022}. However, conclusive results on the emergence of non-trivial topological states are still lacking to date, although there has been some recent progress \cite{Dvir2023,Haaf2024,Zatelli2024}.

In parallel, hybrid devices such as Josephson junctions (JJs) with a strong spin-orbit material as the normal channel constitute an interesting platform for non-reciprocal dissipation-less transport~\cite{Turini2022, Davydova_2022, Souto_2022, Baumgartner2022, Wang_2024, Pillet_2023,Coraiola2024,Fracassi2024}. Indeed, in the presence of a finite magnetic field, the so-called Josephson diode effect has been demonstrated~\cite{Turini2022, Baumgartner2022}, which is related to the simultaneous breaking of both inversion and time-reversal symmetry. These achievements opened interesting possibilities for future applications in low-power superconducting electronics. In this respect it is thus important to inspect tools such as electrostatic gates that allow for easy and precise manipulation of the device performance~\cite{Guiducci2019a, Guiducci2019, Iorio2019, Casparis2019, Beukman2017, Elalaily2021, KovacsKrausz2020}.

InSb nanoflags represent a promising semiconductor material platform, due to their strong intrinsic spin-orbit interactions and quasi-2D character~\cite{Moehle2021,Verma2021,Ke2019,Vries2019,Salimian2021,Sladek1957,Qu2016,Mata2016,Chen2021a,Vurgaftman2001,Lei2021}. Previous studies have reported detailed characterization of planar JJs based on InSb nanoflags~\cite{Vries2019,Salimian2021,Turini2022,Iorio2023}, showing ballistic transport and sizeable Josephson diode effect~\cite{Turini2022}.

In this work, we consider InSb nanoflag-based JJs fabricated with lateral electrostatic gates (side-gates). The aim is to demonstrate that side-gates can be used as a tool to manipulate device transport properties both in the dissipative and in the dissipation-less regime.

\section{Results}
\subsection{Device geometry}
\label{ch:magnetotransport}

\begin{figure}[ht]
    \centering
    \includegraphics[width=\linewidth]{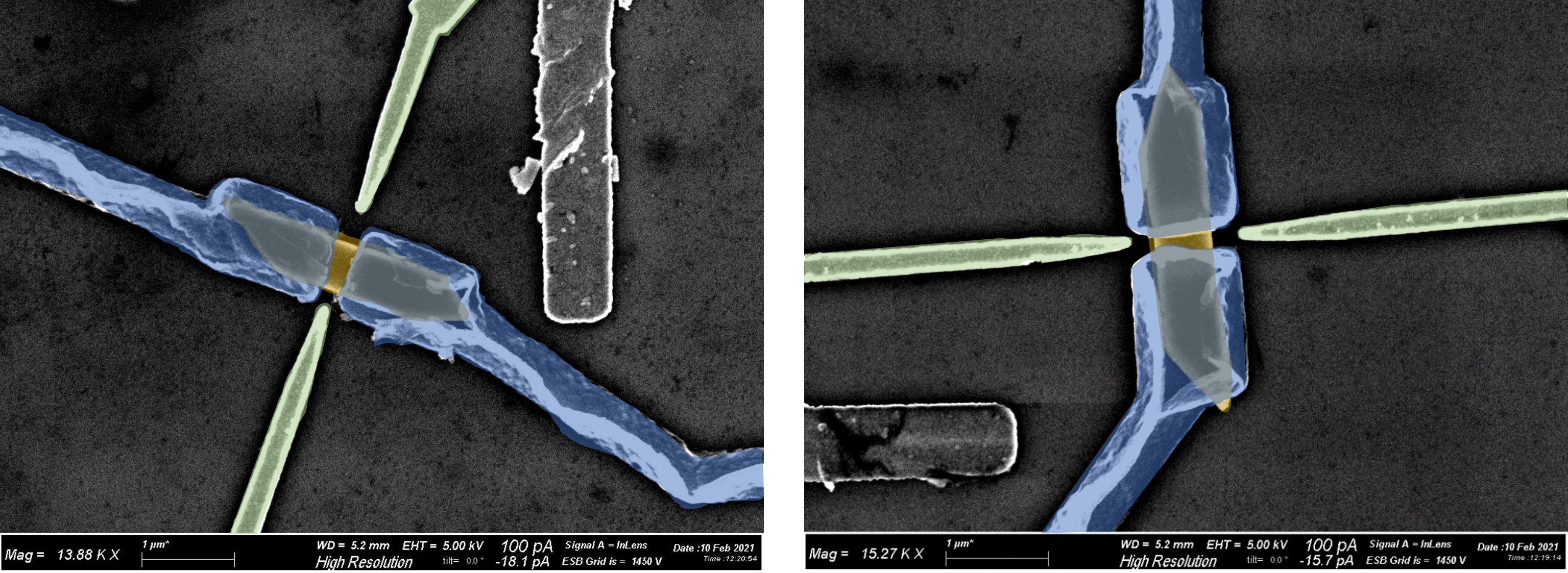}
    \caption{Device F6 and Device F7 from sample SC20. The SEM images have been color-coded: InSb orange, Nb blue, and side-gates green.}
    \label{fig:SC20}
\end{figure}

In this work, two devices have been studied: SC20-F6 and SC20-F7. The devices, in addition to a conventional back-gate, are fabricated with two lateral gates, or \textit{side-gates}, which are placed at a distance of about $250$~nm from the nanoflag, as shown in Figure \ref{fig:SC20} (see the Experimental Section for details on the device fabrication). When the side-gates are grounded, the devices behave consistently with what reported in Refs.~\cite{Salimian2021,Turini2022,Iorio2023}, where InSb nanoflag-based JJs without side-gates have been investigated.

Here we concentrate on the impact of side-gates on transport properties and supercurrent flow, while for a detailed characterization of JJ behavior, e.g., back-gate modulation and non-reciprocal supercurrent transport, we refer the reader to Refs.~\cite{Salimian2021,Turini2022,Iorio2023} and to Section S1 in the Supplementary Information.

\subsection{Conductance modulation by the gates}
\label{sec:ch4-cond_mod}

\begin{figure}[ht]
    \centering
    \includegraphics[width=\linewidth]{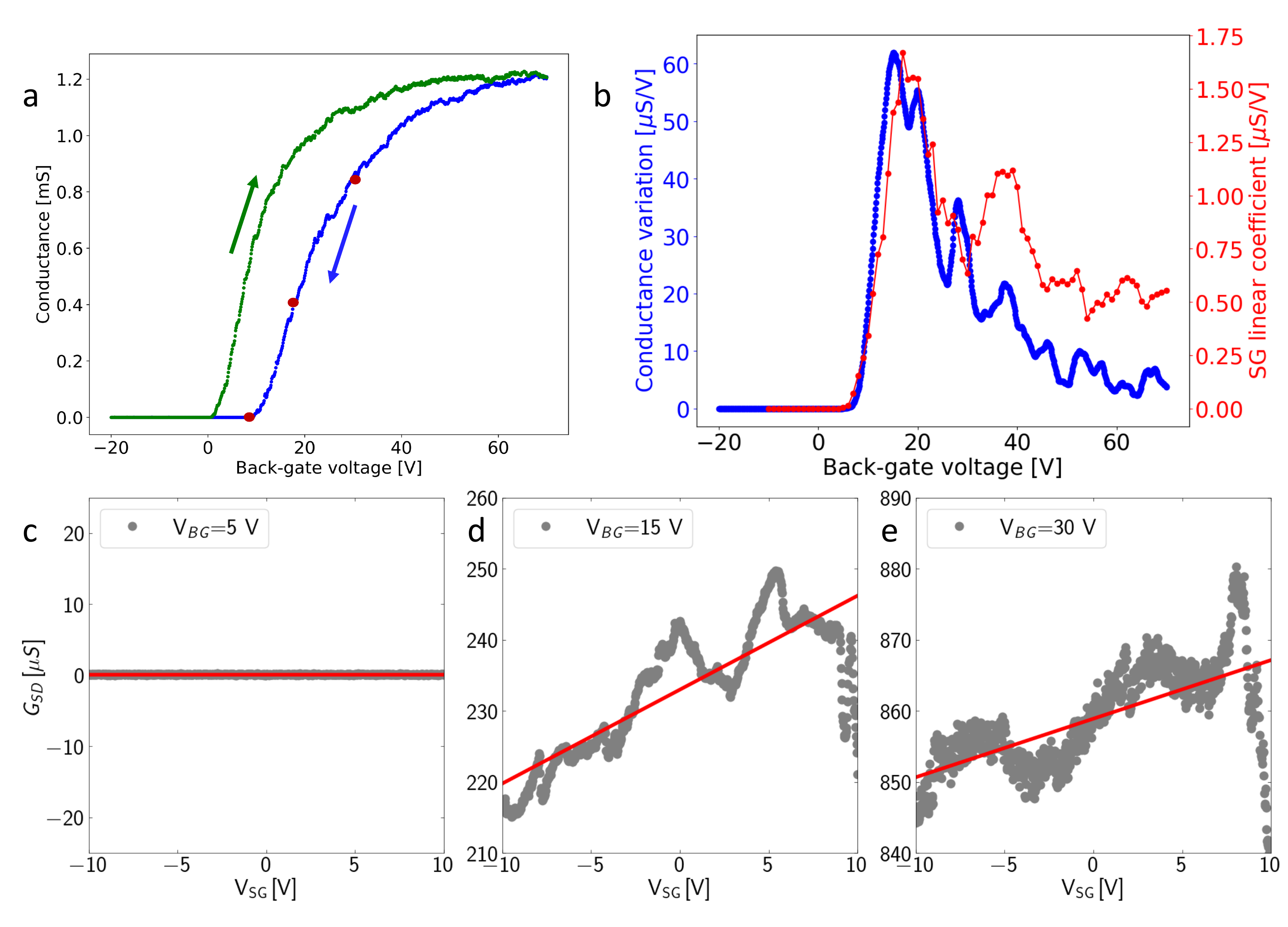}
    \caption{Back-gate and side-gate modulation of the conductance. (a) Source-drain conductance as a function of back-gate voltage. The arrows indicate the sweep direction. The three red dots indicate the back-gate values for which the data in (c-e) were taken. (b) Comparison between the conductance variation with respect to the back-gate voltage (left axis) and the side-gate voltage (right axis). The two quantities have analogous structures but they differ in magnitude by a factor $30$. (c-e) Conductance versus side-gate voltage for $V_{BG}$ $=$ $5$~V, $15$~V, and $30$~V, respectively. Grey points represent the experimental data, while the red lines are linear fitting curves. $T$ $=$ $4.2$ K. $B$ $=$ $0$. Device SC20-F6.}
    \label{fig:ch4-G6vsG}
\end{figure}

The main focus of this section is to quantify the efficiency of the side-gates, with respect to the global back-gate, and to extract valuable information about the charge dynamics in the nanoflag, via magneto-transport measurements. For these measurements, the devices are kept in the normal state, since the chosen bias current is higher than the critical current $I_c$ (see Section S1.1 of the SI). Firstly, the device conductance is measured as a function of back-gate voltage ($V_{BG})$, at $T=4.2$ K, as shown in Figure \ref{fig:ch4-G6vsG}(a) for device SC20-F6. The acquisitions are performed with a low frequency AC current bias set-up. We observe that the pinch-off value of the device, i.e., the value of back-gate voltage below which the conductance is negligible, depends on the sweep direction. In the forward direction, the conductance is zero below $\sim$$1$~V, while, in the backward direction, it drops to almost zero below $\sim$$7$~V. Hysteresis is known to appear in this kind of devices because of defect-assisted charge trapping \cite{Nanoflags2021}. To have a consistent set of data, in the following, each sweep in back-gate voltage is performed starting from the high-conductance plateau; each time that a higher voltage is needed, the back-gate is brought to $V_{BG} = 70$~V and then decreased to the desired value. A measurable conductance value is found only for positive $V_{BG}$, which indicates that the semiconductor majority carriers are electrons. Thus, the InSb nanostructure is of n-type.

				The same measurement is repeated in presence of a non-zero side-gate voltage $V_{SG}$. Using the symmetric configuration ($V_{SG1} = V_{SG2} = V_{SG}$), the conductance is measured as a function of $V_{SG}$, at fixed values of $V_{BG}$. Three traces are shown in Figure \ref{fig:ch4-G6vsG}(c-e) at the working points $V_{BG} = 30$~V, $15$~V, and $5$~V, respectively. A linear fit is performed for each curve, from which the value of the slope $m(V_{BG})$ is extracted. Figure \ref{fig:ch4-G6vsG}(b) shows the dependence of this slope versus back-gate voltage working point, compared with the differential variation of the conductance with respect to the back-gate $dG/dV_{BG}$. The latter is calculated via numerical differentiation of the curve in Figure \ref{fig:ch4-G6vsG}(a). The two curves have similar behavior: in both cases, the maximal modulation occurs for $V_{BG} \sim 15$~V, and no change is induced below the pinch-off value $7$~V. Note that the two scales differ by a factor $30$. From this analysis, we deduce that the relative efficiency of the side-gates is $1/30$.

Individual side-gate sweeps show a background modulation which can be related to universal conductance fluctuations (UCF). This is a common feature of mesoscopic systems~\cite{Hansen2005, EstevezHernandez2010, Roulleau2010, Iorio2019}, in which the phase coherence length is larger than the geometrical size. Based on the random matrix theory \cite{Beenakker1997}, a fluctuation of magnitude $G_0=2e^2/h$, i.e., one conductance quantum, is expected in structures which do not show any spin texture. Here, the observed fluctuations have a magnitude of the order of $10$~$\mu$S, corresponding to $\sim0.3$~$G_0$. This reduced value is consistent with predictions for a system with a strong Rashba spin-orbit interaction \cite{Choe2015}.

\begin{figure}[ht]
    \centering
    \includegraphics[width=0.7\linewidth]{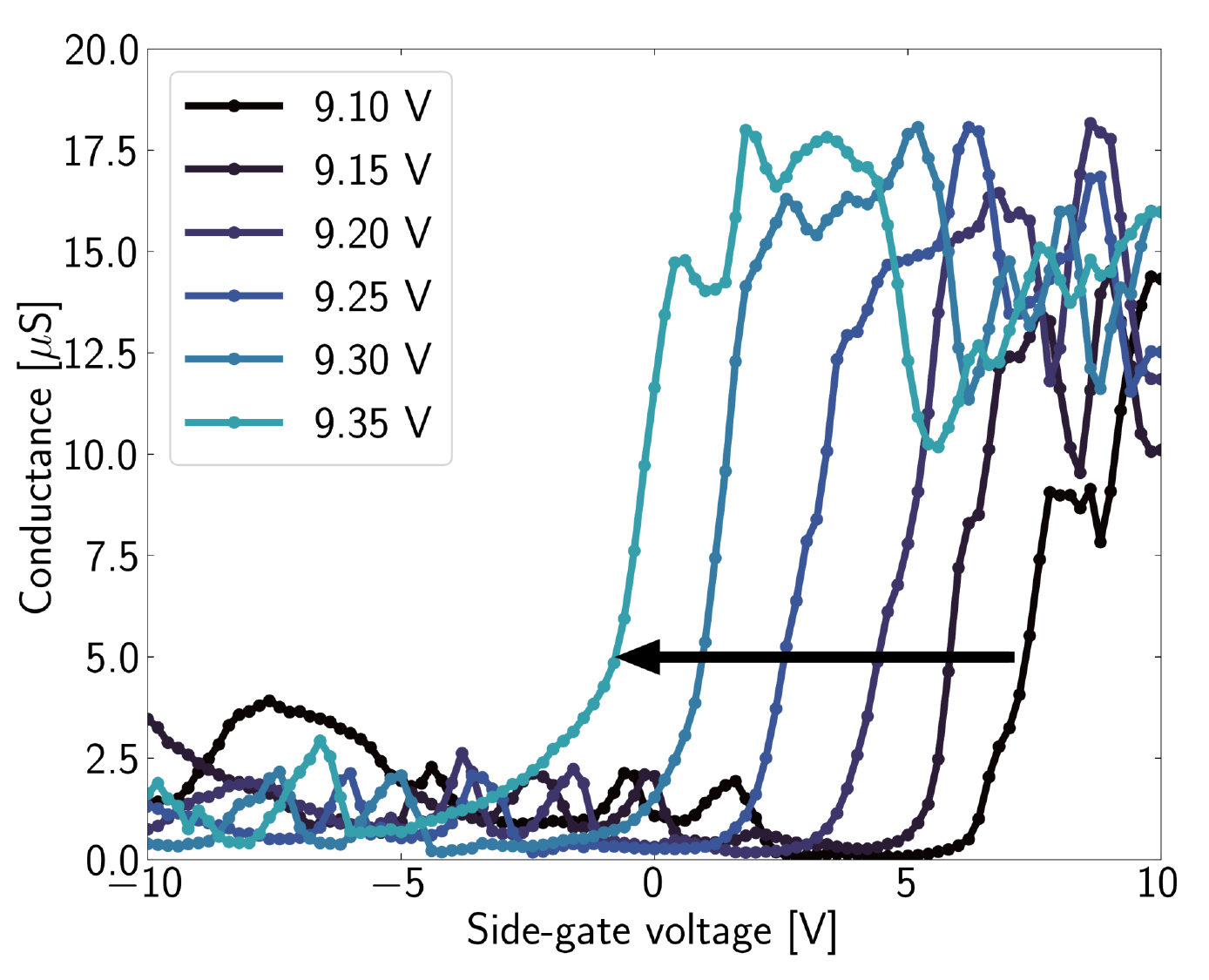}
    \caption{Conductance pinch-off versus side-gate voltage. Six traces are reported, for different values of $V_{BG}$ between $9.1$~V and $9.35$~V. The conductance drops for $V_{SG}$ below a threshold voltage, which depends monotonically on $V_{BG}$. $T$ $=$ $250$~mK. $B$ $=$ $0$. Device SC20-F6.}
    \label{fig:pinch-off}
\end{figure}

Finally, we report that the side-gates efficiently pinch off the device, if the working back-gate point is properly chosen. Figure \ref{fig:pinch-off} presents conductance traces \textit{versus} side-gate voltage, for different values of $V_{BG}$, close to the pinch-off. Conductance drops to zero below a threshold value, which depends on the working point in $V_{BG}$.

\subsection{Magnetic field-induced interference pattern}
\label{sec:Fraunhofer}

We now consider the effect of side-gates when supercurrent flows through the JJ. The supercurrent flow depends on the magnetic flux through the normal region of the JJ \cite{Tinkham2004}. Due to this mechanism, it is possible to extract information about the effective geometry of the system by looking at the dependence of the $I-V$ curves as a function of applied perpendicular magnetic field. In this configuration, where the magnetic field vector $\mathbf{B} = B \hat{z}$ is aligned with the normal direction of the junction plane ($x-y$), the magnetic flux is
\begin{equation}
    \Phi(B) = A_{eff}B_{eff} = W(L+2\lambda_L)FB,
\end{equation}
with $L$ and $W$ the length and width of the JJ, respectively. Here, the effective area $A_{eff}$ is not directly given by the geometric dimensions of the junction, since the magnetic field penetrates the two superconducting Niobium leads for a distance defined by the London penetration length ($\lambda_L)$. This quantity only depends on the superconducting material and equals $\lambda_L=40$~nm in Niobium \cite{Maxfield1965}. Moreover, the magnetic field lines are deflected in presence of the superconducting leads, due to the Meissner effect \cite{Tinkham2004, Barone1982, Meissner1933}, and therefore the effective value of the magnetic field is increased. This effect can be taken into account with the flux focusing coefficient $F$, such that $B_{eff} = FB$ \cite{Guiducci2019,Casparis2019,Paajaste2015,Rosenthal1991,Kjaergaard2016}.

\begin{figure}[ht]
    \centering
    \includegraphics[width=\linewidth]{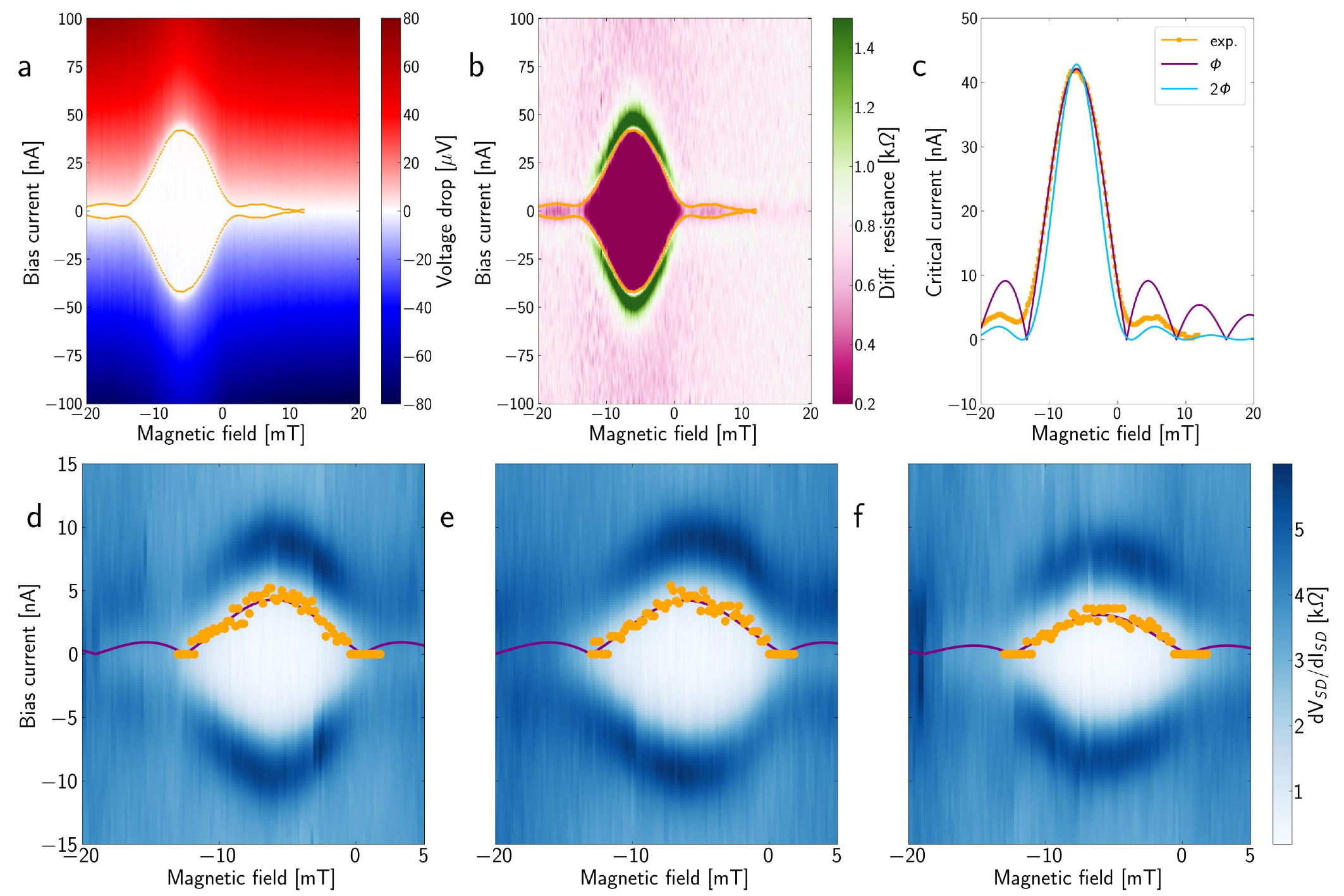}
    \caption{(a-c) Interference pattern of the critical current in presence of a perpendicular magnetic field. (a) Voltage drop and (b) differential resistance, obtained by numerical differentiation, are reported as a function of perpendicular magnetic field ($x$-axis) and bias current ($y$-axis). The critical current is indicated by orange dots. (c) Comparison between the extracted values of critical current $I_c$ and the models in Eq.~(\ref{eq:fit-1-F}) (purple) and Eq.~(\ref{eq:fit-2-F}) (light blue). A robust superconducting region is present, centered at $B$ $=$ $-6$~mT, due to a residual magnetization of the cryostat (see text). The first side lobes weakly appear in the experimental data. For the measurements shown in (a-c), the side-gates were grounded. (d-f) Fraunhofer pattern for different values of the side-gate voltage $V_{SG}$. (d) $V_{SG} = 10$~V, (e) $V_{SG} = 0$~V, and (f) $V_{SG} = -10$~V. The two-dimensional maps show the differential resistance \textit{versus} perpendicular magnetic field and bias current. The orange dots represent the experimental values of $I_c$, while the purple lines are the best-fit curves based on the model in Eq.~(\ref{eq:fit-1-F}). (a-c) $V_{BG} = 40$~V, and (d-f) $V_{BG} = 16$~V. $T = 250$~mK. Device SC20-F6.} 
    \label{fig:Fraunhofer}
\end{figure}

Figure \ref{fig:Fraunhofer}(a,b) shows the $I-V$ characteristics and the differential resistance, respectively, as a function of bias current and magnetic field $B$, obtained at $V_{BG} = 40$ V and $V_{SG} = 0$ for device SC20-F6. The value of the critical currents $I_c(B)$ is extracted by setting a threshold voltage of $V_{th} = 2$~$\mu$V in the $I-V$ curves, and is indicated by orange dots in Figure \ref{fig:Fraunhofer}. A robust dissipation-less region is present for $-12$~mT $\le$ B $\le$ $0$~mT: the lobe is centered in $B_0 = -6$~mT, and its half-width is $\Delta B=7.3$ mT. The maximal supercurrent in a single planar junction is, in fact, expected for $B=0$. We attribute the small offset to a residual magnetization of the cryostat.

Observation of a Fraunhofer pattern is expected in a planar Josephson junction, under the assumption of (i) a uniform supercurrent distribution $\mathbf{J}(x)$,  and (ii) a wide junction ($W>>L$). However, if $W/L$ is finite, the effective periodicity is larger than the ideal prediction \cite{Heida1998} of the superconducting flux quantum $\Phi_0=h/(2e)$. Indeed, the occurrence of an anomalous Fraunhofer pattern has been previously associated with the geometry of the system \cite{Heida1998}. In Ref.~\cite{Barzykin1999}, a continuous and non-monotonic change of the periodicity from $\Phi_0$ to $2\Phi_0$ is demonstrated when the ratio $L/W$ grows.

Our experimental data show only a weak trace of the first lateral lobes. Notably, the height of these features is of the order of the intrinsic
thermal current noise of the junction $\delta I_{th} = 2 e k_B T / \hbar$ \cite{Likharev1979,Zhi2019}; here, $\delta I_{th} = 10.5$ nA. Therefore thermal fluctuations may mask the side lobes. Nevertheless, a trace of the first lateral lobes is visible in the differential resistance.

The devices under study present a ratio $(L+2\lambda_L)/W\sim0.5$. The orange (light blue) line in Figure \ref{fig:Fraunhofer}(c) stands for the best fit function in the limiting case $L/W<<1$ ($L/W>>1$). The two models are \cite{Heida1998,Barzykin1999,Mur1996,Dubos2001}
\begin{eqnarray}
    I_c & = & I_{c,0}\left|\frac{\sin\bigg(\frac{\pi\Phi(B)}{\Phi_0}\bigg)}{\frac{\pi\Phi(B)}{\Phi_0}}\right| \quad L/W<<1 \label{eq:fit-1-F},\\
    I_c & = & I_{c,0}\left[\frac{\sin\bigg(\frac{\pi\Phi(B)}{2\Phi_0}\bigg)}{\frac{\pi\Phi(B)}{2\Phi_0}}\right]^2 \quad L/W>>1 \label{eq:fit-2-F},
\end{eqnarray}
respectively, with $\Phi(B) = A_{eff} F(B - B_0)$. The center of the interference pattern $B_0$, the critical current for $B = B_0$, i.e., $I_{c,0}$, and the flux focusing factor $F$ are left as free parameters. When Eq.~(\ref{eq:fit-1-F}) is used, the maximal critical current $I_{c,0}=42$ nA is consistent with the experimental observation. The resulting flux focusing coefficient $F = 1.6$ is comparable to results found in literature \cite{Paajaste2015}. On the other hand, the model in Eq.~(\ref{eq:fit-2-F}) results in $F = 2.3$. As shown in Figure \ref{fig:Fraunhofer}(c), Eq.~(\ref{eq:fit-1-F}), represented by the purple line, fits well the central region, while the side lobes are better modeled by Eq.~(\ref{eq:fit-2-F}) (light blue line). This intermediate regime is consistent with the ratio $L/W$ of the devices.

The Fraunhofer pattern measurement was repeated for different values of $V_{SG}$ of the side-gates in symmetric configuration. The Fraunhofer patterns obtained at $V_{BG} = 16$ V for $V_{SG} = 10$~V, $0$~V, and $-10$~V are presented in Figure \ref{fig:Fraunhofer}(d-e-f), respectively. In each plot, the experimental values of $I_c$ are shown as orange dots, while the purple line is the best fit curve obtained with the model in Eq.~(\ref{eq:fit-1-F}).

By using the flux focusing factor $F = 1.6$ previously estimated, we can compare the effective area $A_{eff}$ for each configuration. The width of the fitting function ($w$), which is inversely dependent on $A_{eff}$, changes for the three configurations. The values are reported in Table \ref{tab:sidegates}. The monotonic trend is in line with the expected result: when the side-gate voltage is negative, the charge density in the nanoflag is locally reduced in the region where $V_{SG}$ is more effective, i.e., the external part of the junction. This implies that the current density $\mathbf{J}(x)$ is spatially-dependent. This can be seen as an effective reduction of the channel width, which would increase the width of the Fraunhofer pattern, consistent with our results. At the same time, the critical current $I_{c,0}$ at $B = B_0$ is reduced by decreasing $V_{SG}$, in a similar fashion to the back-gate dependence. In summary, we have shown that the presence of side-gates can impact the supercurrent flow through the JJ, modulating both the critical current amplitude of the interference pattern and the effective area of the junction.

\begin{table}[ht]
    \caption{Results of the fit procedure of the Fraunhofer pattern for different $V_{SG}$. $T = 250$~mK. $V_{BG} = 16$~V. Device SC20-F6.}
    \begin{tabular}{c|ccc}
		  \hline
      $V_{SG}$ [V] & 10&0&-10 \\
      \hline
      $w$ [mT] & $7\pm1$ & $8\pm1$ & $9\pm1$\\
      $I_{c,0}$ [nA] & $5.4\pm0.2$&$5.0\pm0.2$&$4.0\pm0.2$\\
			\hline
    \end{tabular}
    \label{tab:sidegates}
\end{table}

\subsection{Conductance behavior at high magnetic fields}
\label{sec:ch4-cond_quant}

\begin{figure}[t]
    \centering
    \includegraphics[width=\linewidth]{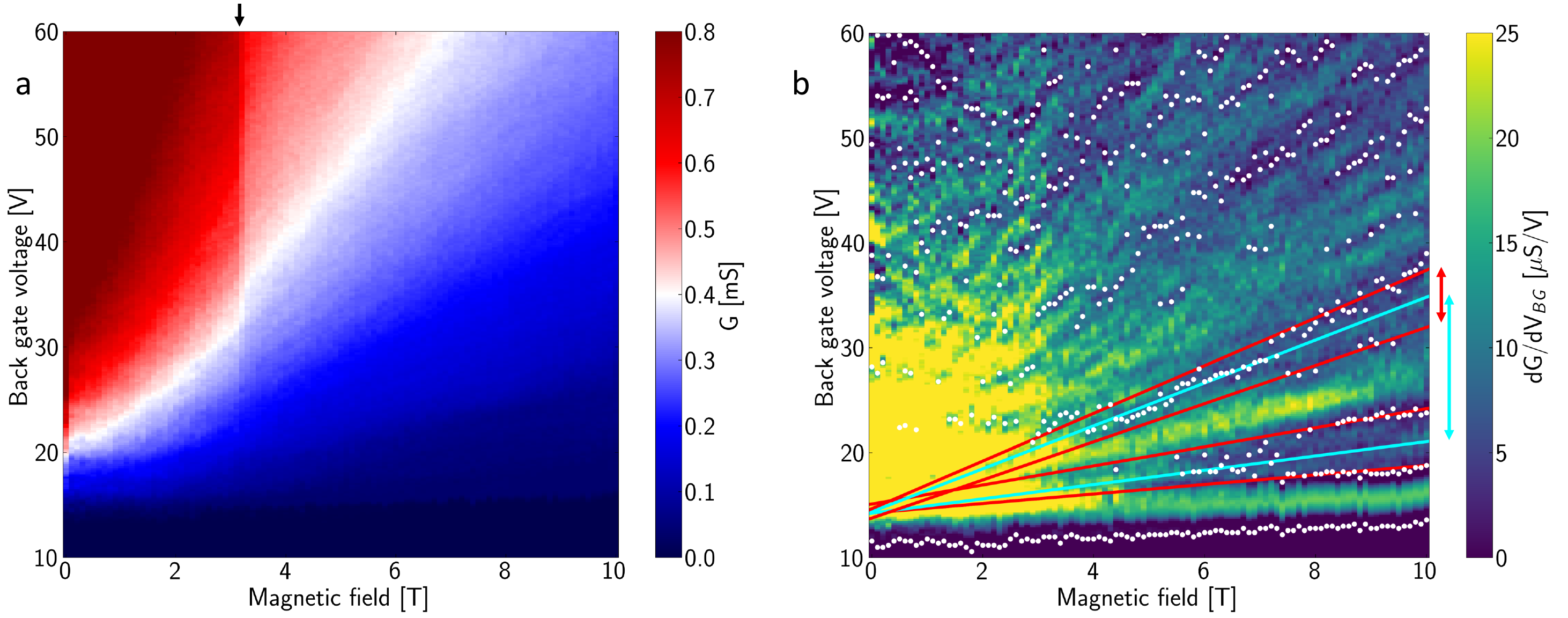}
    \caption{Conductance modulation as a function of perpendicular magnetic field and back-gate voltage. (a) Conductance as a function of $B$ ($x$-axis) and $V_{BG}$ ($y$-axis). A kink is present at $B$ $=$ $3.15$ T, indicated by the black arrow, due to the transition of the contacts (see text). (b) Conductance variation with respect to $V_{BG}$, versus $B$ ($x$-axis) and $V_{BG}$ ($y$-axis). The white dots represent the local minima of the traces. Two families of minima follow the light-blue lines for $B$ $\leq$ $7$~T, and subsequently spit up in 4 total traces, indicated by the red lines. The arrows denote the line separation at $B$ $=$ $10$ T. $T$ $=$ $250$~mK. Device SC20-F7.}
    \label{fig:ch4-landaufan}
\end{figure}

Next, we inspect the behavior of the device when a strong perpendicular magnetic field is applied, with the two side-gates grounded ($V_{SG} = 0$). At high magnetic fields, conductance is predicted to have quantized values with respect to back-gate voltage, due to the formation of Landau levels. The interplay between magnetic field and carrier density modulation is shown in Figure \ref{fig:ch4-landaufan}(a) for device SC20-F7, which represents conductance as a function of a perpendicular magnetic field ($x$-axis) and back-gate voltage ($y$-axis). Here, we note a kink at $B=3.15$ T, which corresponds to the transition of the Nb contacts from the superconducting to the normal state. To observe magneto-conductance effects, the conductance variation $dG/dV_{BG}$ is shown in Figure \ref{fig:ch4-landaufan}(b).

In Figure \ref{fig:ch4-landaufan}(b), the relative minima are clearly collected in straight lines. We obtained the relative minima for each trace; the resulting points are shown as white dots. For 4 T $< B <$ 7 T, two families of minima are present, whose slope is $m_0 = 0.7 \pm 0.1$~V/T and $m_1 = 2.1 \pm 0.1$~V/T. These are represented by orange lines in Figure \ref{fig:ch4-landaufan}(b). For higher values of the magnetic field ($B > 7$~T), each of the two curves splits up into two (pink lines), symmetrically with respect to the original lines.
 
Following the Shubnikov-de-Haas theory of magneto-transport in two-dimensional electron gases, the conductance is predicted to show a set of minima located on straight lines, emerging from integer filling  of Landau levels (LLs). The energy spectrum $E_n$ is discrete, $E_n = \hbar \omega_c (n + \frac{1}{2})$, where $\omega_c = eB/m^*$ is the cyclotron frequency, $\hbar$ the reduced Planck constant, and $m^*$ the effective mass \cite{Ihn2009}, i.e., the LL energy depends linearly on the magnetic field. The two orange lines indicate the two lowest spin-degenerate Landau levels ($n=0$ and $n=1$), corresponding to filling factors $\nu=2$ and $\nu=4$, respectively. Introducing the back-gate lever arm $\alpha_{BG}$ as the ratio between induced charge $\rho_{2D}$ and applied back-gate voltage $V_{BG}$, $\alpha_{BG} = \rho_{2D} / V_{BG}$, the slope of the curves is related to the quantum number $n$ as
\begin{equation}
m_n= \frac{\hbar e}{m^*}\frac{\rho_{2D}}{\alpha_{BG}} \left(n+\frac{1}{2}\right),
\end{equation}
where $\rho_{2D}=m^*/\pi\hbar^2$ is the spin-degenerate density of states in the semiconductor per unit area and $e$ the elementary charge. The ratio between the first two coefficients $m_1/m_0=3$ is consistent with the experimental evidence. Moreover, due to the large effective g-factor of InSb, the levels are expected to be spin-resolved for high fields, with a gap given by the Zeeman energy. In our case, the relevant energy scales are the spacing between Landau levels $E_L$ and the Zeeman energy $E_Z$
\begin{eqnarray}
     E_L & = & \hbar\omega_c = \frac{\hbar eB}{m^*}\\
     E_Z & = & g^*\mu_BB,
\end{eqnarray}
where $g^*$ is the effective g-factor associated to spin-orbit interactions and $\mu_B$ the Bohr magneton. The separation between the two orange lines at $B=10$ T, in units of back-gate voltage, is $\delta E_L = 13.9$~V and $\delta E_Z = 5.5$~V for the original (orange) and split (pink) lines, respectively. These values are indicated by the arrows in Figure \ref{fig:ch4-landaufan}(b). By comparing them, we can extract the effective g-factor, resulting in
\begin{equation}
 g^* = \frac{\delta E_Z}{\delta E_L}\frac{\hbar e}{m^* \mu_B} = 44,
\end{equation}
which is consistent with the values reported in literature for InSb nanostructures \cite{Litvienko2008}.

\section{Conclusions}

We have reported measurements on InSb nanoflag-based JJ devices where two side-gates have been laterally placed $250$~nm from the junction.
We have shown that side-gates can completely pinch-off the device, similarly to the back-gate action, with a relative efficiency of $1/30$. In the dissipation-less regime, both critical current amplitude and Fraunhofer interference pattern can be manipulated by means of the side-gates. Thus, side-gates represent an useful additional electrical knob to be used in conjunction to back-gates~\cite{Guiducci2019}, and we expect that their efficiency can be further improved by fabricating them closer to the flag. The electrostatic control of InSb nanoflags with side-gates might also be used to tune spin-orbit interactions~\cite{Iorio2019, KovacsKrausz2020}, which could result in an additional knob for the Josephson diode effect. 

The observation of Landau levels at high magnetic fields clearly shows that the transport in the nanoflags has a two-dimensional character, while the measured $g^*$ confirms the strong spin-orbit interaction predicted for InSb-based devices.

\section{Experimental Section}
\label{ch:devices}

The InSb nanoflags were grown by chemical beam epitaxy (CBE) from metal-organic precursors in a Riber Compact-21 system. Tapered Indium Phosphide nanowires (NW) are used to provide full support to the nanoflags. This strategy allowed to obtain InSb nanoflags of $(2.8\pm0.2)\;\mu$m length, $(470\pm80)$ nm width, and $(105\pm20)$ nm thickness \cite{Verma2021,Verma2020}.

Initially, InP nanowires were grown on InP(111)B substrates via Au-assisted growth. The catalyst particles were 30 nm Au colloids dropcasted onto the bare substrate. The InP nanowire stems were grown with sample rotation for 90 minutes at $400^{\circ}$C, with 0.6 Torr TMIn (trimethylindium) and 1.2 Torr TBP (tert-butylphosphine). Then the growth temperature was reduced by 30$^{\circ}$C (in presence of TBP flux only) to the InSb growth temperature. The InSb nanoflags were grown without rotation, after aligning the $\langle112\rangle$ crystal direction toward the Sb precursor beam, as described in \cite{Verma2021,Verma2020}, initially for 30 min with 0.6 Torr TMIn and 2.3 Torr TMSb (trimethylantimony), and then for additional 60 min, linearly increasing the TMSb line pressure from 2.3 to 2.6 Torr, in order to enhance the asymmetric radial growth.

The Josephson junctions were defined by deposition of Niobium contacts. Figure \ref{fig:SC20}(a,b) reports images of devices SC20-F6 and SC20-F7, respectively, captured by scanning electron microscopy (SEM). These devices are equipped with two side-gates, made in Cr/Au as described in the following.

Firstly, the InSb nanoflags are dry transferred onto a pre-patterned highly conductive p-type Si(100) substrate, which serves as a global back-gate. A $285$ nm thick SiO$_2$ layer covers the Si substrate as dielectric. During the mechanical transfer, the InSb nanoflags are detached from the InP NW stems, resulting in well-isolated InSb nanoflags, which were randomly distributed on the substrate. The position of selected InSb nanoflags was determined relative to predefined alignment markers using SEM images. Considering the thickness and the edge geometry of the InSb nanoflags, electrodes were patterned on a $400$ nm thick layer of AR-P $679.04$ resist with standard electron-beam lithography (EBL). Prior to metal deposition, the native oxide of the InSb was removed using a sulfur-based etching solution which results also in a smoother InSb surface \cite{Guel2017}. To this end, the native oxide was etched for 30 s in an optimized sulfur solution of (NH$_4$)$_2$S$_x$ (1:9 (NH$_4$)$_2$S$_x$:DI water at $40^{\circ}$ C). Then the samples were rinsed in DI water for $30$ s. Next, a $10/150$ nm Ti/Nb film was deposited with high deposition rate after an intense pre-sputtering of each target, followed by lift off in hot acetone. In order to fabricate side-gates, another lithographic step was performed, using AR-P 679.04 resist. In this step, 10 nm/130 nm of Cr/Au were deposited by thermal evaporation.

Table \ref{tab:dimensions} collects the geometric dimensions of the nanoflag-based devices, measured from SEM images.

\begin{table}[t]
    \caption{Device dimensions extracted from SEM images. $W$ stands for the nanoflag width, $L$ is the average normal region length, $d$ is the distance of the side-gates from the nanoflag, computed as the mean between the left-side and right-side-gate. The error is $\pm10$ nm.}
    \begin{tabular}{c|ccc}
		  \hline
      Device    & $W$ [nm] & $L$ [nm]&$d$ [nm]\\
      \hline
      SC20-F6  &$540$   &$250$  &$260$\\
      SC20-F7  &$530$   &$110$  &$240$\\
		  \hline
    \end{tabular}
    \label{tab:dimensions}
\end{table}

While device SC20-F7 was measured in a \textit{quasi-four} terminal configuration, in SC20-F6, three-probe measurements were performed, because of a damage in one lead. This resulted in a series resistance of $R_s= 2500\;\Omega$, measured at $T=4$ K, which was numerically subtracted before processing the data.

For the measurements, we have used an \textit{Oxford Heliox} $^3$He cryostat, whose base temperature is $T = 250$~mK. The cryostat is equipped with a magnet whose field is perpendicular to the junction plane. The insert is equipped with RC- and $\pi$-filters. For AC measurements, a lock-in amplifier (LIA) was used to generate a sinusoidal voltage, with frequency $f=13.321$ Hz, which was sent to the sample through a $10$ M$\Omega$ resistance. By measuring both current and voltage drop across the junction with phase-sensitive detection, the conductance of the device can be properly evaluated.

For measurements in DC configuration, in the same $^3$He cryostat, current bias configuration was used. To improve the signal-to-noise ratio, the sampled signals in DC configuration pass through low-noise pre-amplifiers before being measured.

\begin{acknowledgments}
The authors acknowledge the support from the project PRIN2022 2022-PH852L(PE3) TopoFlags - ``Non reciprocal supercurrent and topological transition in hybrid Nb-InSb nanoflags'' funded by the European community - Next Generation EU within the programme ``PNRR Missione 4 - Componente 2 - Investimento 1.1 Fondo per il Programma Nazionale di Ricerca e Progetti di Rilevante Interesse Nazionale (PRIN)'' and by PNRR MUR Project No. PE0000023-NQSTI.
\end{acknowledgments}

\bibliography{bianca-jde}

\end{document}